\documentclass[a4paper,11pt]{article}
\usepackage{pos}

\newcommand{\Pom}{\mathbb{P}}

\newcommand{\p}{\partial}

\newcommand{\twosidep}[1]{\stackrel{\leftrightarrow}{\p}_{\! #1}}

\newcommand*\wideestimates{\mathrel{\widehat{=}}}

\title{Central exclusive production of axial-vector $f_{1}$ mesons in proton-proton collisions}

\author[a]{P. Lebiedowicz}
\author[b]{J. Leutgeb}
\author*[c]{O. Nachtmann}
\author[b]{A. Rebhan}
\author[a,\dag]{A. Szczurek}
\notes{\note{Also at \textit{College of Natural Sciences, 
Institute of Physics, University of Rzesz{\'o}w, 
Pigonia 1, PL-35310 Rzesz{\'o}w, Poland}.}}
\affiliation[a]{Institute of Nuclear Physics Polish Academy of Sciences,\\ 
Radzikowskiego 152, PL-31342 Krak{\'o}w, Poland}

\affiliation[b]{Institut f\"ur Theoretische Physik, Technische Universit\"at Wien,\\
Wiedner Hauptstrasse 8-10, A-1040 Vienna, Austria}

\affiliation[c]{Institut f\"ur Theoretische Physik, Universit\"at Heidelberg,\\
Philosophenweg 16, D-69120 Heidelberg, Germany}

\emailAdd{Piotr.Lebiedowicz@ifj.edu.pl}
\emailAdd{josef.leutgeb@tuwien.ac.at}
\emailAdd{O.Nachtmann@thphys.uni-heidelberg.de}
\emailAdd{anton.rebhan@tuwien.ac.at}
\emailAdd{Antoni.Szczurek@ifj.edu.pl}

\abstract{The production of $f_{1}$ ($J^{PC} = 1^{++}$) mesons
in proton-proton collisions via pomeron-pomeron fusion is discussed.
Two ways to construct the pomeron-pomeron-$f_{1}$ coupling
are presented. Comparisons with data from the WA102 experiment are made
and predictions for RHIC and LHC experiments are given.}

\FullConference{%
  40th International Conference on High Energy physics - ICHEP2020\\
  July 28 - August 6, 2020\\
  Prague, Czech Republic (virtual meeting)
}

\begin{document}
\maketitle

\section{Introduction}
\vspace{-0.3cm}

\begin{figure}[!h]
\center
\includegraphics[width=.35\textwidth]{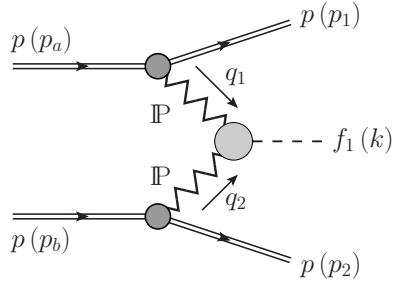}
\caption{Diagram for the reaction (\ref{1.1}) 
with double-pomeron exchange (i.e., $\Pom \Pom$-fusion mechanism).}
\label{fig1}
\end{figure}

In this contribution we will be concerned with central exclusive
production (CEP) of $f_{1}(1285)$ and $f_{1}(1420)$ mesons
in proton-proton collisions
\begin{equation}
p\,(p_{a}) + p\,(p_{b}) \to p\,(p_{1}) + f_{1}\,(k) + p\,(p_{2}) \,.
\label{1.1}
\end{equation}
The presentation is based on \cite{Lebiedowicz:2020yre} where
all details and many more results can be found.
At high energies the reaction (\ref{1.1}) should be mainly
due to double-pomeron exchange (figure~\ref{fig1}).

The relevant kinematic quantities are
\begin{eqnarray}
&&s = (p_{a} + p_{b})^{2} 
\qquad {\rm c.m. \; energy\; squared}, \nonumber \\
&&q_1 = p_{a} - p_{1}, \quad q_2 = p_{b} - p_{2}, \quad k = q_{1} + q_{2}, \nonumber \\
&&t_1 = q_{1}^{2}, \quad t_2 = q_{2}^{2}, \quad m_{f_{1}}^{2} = k^{2}\,.
\label{1.2}
\end{eqnarray}

We treat our reaction in the \underline{tensor-pomeron approach}
as introduced in \cite{Ewerz:2013kda}.
This approach has a good basis from nonperturbative QCD
using functional integral techniques \cite{Nachtmann:1991ua}.
We describe the pomeron and the charge-conjugation $C = +1$
reggeons as effective rank 2 symmetric tensor exchanges,
the odderon and $C = -1$ reggeons as effective vector exchanges.
A tensor character of the pomeron is also preferred
in holographic QCD;
see e.g. \cite{Brower:2006ea,Domokos:2009hm,Iatrakis:2016rvj}.

There are by now many applications of the tensor-pomeron model
to two-body hadronic reactions \cite{Ewerz:2016onn},
to photoproduction,
to DIS structure functions at low $x$, and especially to CEP reactions:
\begin{equation}
p + p \to p + X + p \,, \;\; {\rm where}\;\;
X = \eta,\, \eta',\, f_{0},\, f_{2},\, \pi^{+}\pi^{-},\,
4\pi,\, p \bar{p},\, K\bar{K},\, K\bar{K}K\bar{K},\, \rho^{0},\, \phi,\, \phi \phi\,;
\label{1.3}
\end{equation}
see e.g. \cite{Lebiedowicz:2013ika,Lebiedowicz:2016ioh,Lebiedowicz:2019boz}.

From these works we know the form of the effective $\Pom$ propagator 
and the $\Pom pp$ vertex.
The new quantity in figure~\ref{fig1}, to be studied here,
is the $\Pom \Pom f_{1}$ coupling.

\section{The pomeron-pomeron-$f_{1}$ coupling}
\vspace{-0.3cm}

\begin{figure}[!h]
\center
\includegraphics[width=.26\textwidth]{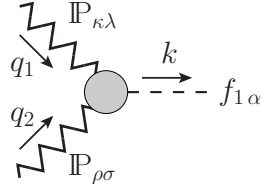}
\caption{Diagram for the $\Pom \Pom f_{1}$ vertex function
$i\Gamma_{\kappa \lambda,\rho \sigma,\alpha}^{(\Pom \Pom f_{1})}(q_{1},q_{2})$ where $q_{1} + q_{2} = k$.}
\label{fig2}
\end{figure}

In this section we describe our ways to construct the Lagrangian
for the $\Pom \Pom f_{1}$ coupling and the corresponding vertex
function (figure~\ref{fig2}).

We follow two strategies for constructing this vertex function.\\

\textbf{(1)} Phenomenological approach.
First we consider a fictitious process: the fusion of two
``real spin two pomerons'' (or tensor glueballs) of mass $m$
giving an $f_{1}$ meson of $J^{PC} = 1^{++}$.
We make an angular momentum analysis of this reaction 
in its c.m. system, the rest system of the $f_{1}$ meson:
\begin{equation}
\Pom\,(m, \epsilon_{1}) + \Pom\,(m, \epsilon_{2}) \to
f_{1}\,(m_{f_{1}},\epsilon)\,.
\label{2.1}
\end{equation}
The spin~2 of these ``pomerons'' can be combined to a total spin $S$
($0 \leqslant S \leqslant 4$) and this must be combined with
the orbital angular momentum $l$ to give the $J^{PC} = 1^{++}$
values of the $f_{1}$.
There are exactly two possibilities for this,
namely
$(l,S) = (2,2)\; \rm{and}\; (4,4)$;
see Appendix~A of \cite{Lebiedowicz:2013ika}.
Corresponding $\Pom \Pom f_{1}$ couplings are easily written down:
\begin{eqnarray}
&
{\cal L}^{(2,2)}_{\Pom \Pom f_{1}} = \frac{g'_{\Pom \Pom f_{1}}}{32\,M_{0}^{2}}
\Big( \Pom_{\kappa \lambda} 
\twosidep{\mu} \,\twosidep{\nu}
\Pom_{\rho \sigma} \Big)
\Big( \p_{\alpha} U_{\beta} - \p_{\beta} U_{\alpha} \Big)\,
\Gamma^{(8)\,\kappa \lambda, \rho \sigma, \mu \nu, \alpha \beta}\,,
\label{2.3}\\
&
{\cal L}^{(4,4)}_{\Pom \Pom f_{1}} = \frac{g''_{\Pom \Pom f_{1}}}{24 \cdot 32 \cdot M_{0}^{4}}
\Big( \Pom_{\kappa \lambda}
\twosidep{\mu_{1}} \, \twosidep{\mu_{2}} \,
      \twosidep{\mu_{3}} \, \twosidep{\mu_{4}}
\Pom_{\rho \sigma} \Big)
\Big( \p_{\alpha} U_{\beta} - \p_{\beta} U_{\alpha} \Big)\,
\Gamma^{(10)\,\kappa \lambda, \rho \sigma, \mu_{1} \mu_{2} \mu_{3} \mu_{4}, \alpha \beta}\,,
\label{2.4}
\end{eqnarray}
where $M_{0} \equiv 1$~GeV (introduced for dimensional reasons),
$\Pom_{\kappa \lambda}$ is the $\Pom$ effective field,
$U_{\alpha}$ is the $f_{1}$ field,
$g'_{\Pom \Pom f_{1}}$ and $g''_{\Pom \Pom f_{1}}$ are
dimensionless coupling constants, and $\Gamma^{(8)}$, $\Gamma^{(10)}$ 
are known tensor functions~\cite{Lebiedowicz:2020yre}.
We use then these couplings, supplemented by suitable
form factors, for the $f_{1}$ CEP reaction.

\textbf{(2)} Our second approach uses holographic QCD, in particular
the Sakai-Sugimoto model \cite{Sakai:2004cn,Brunner:2015oqa}.
There, the $\Pom \Pom f_{1}$ coupling can be derived
from the bulk Chern-Simons term requiring consistency
of supergravity and the gravitational anomaly.
From this we get the following
\begin{eqnarray}
{\cal L}^{\rm CS} &=& \varkappa' \,U_{\alpha}\,\varepsilon^{\alpha \beta \gamma \delta}\,
\Pom^{\mu}_{\;\;\beta}\, \p_{\delta}\Pom_{\gamma \mu}
\nonumber \\
&&+ \;\varkappa'' \,U_{\alpha}\,\varepsilon^{\alpha \beta \gamma \delta}\,
\left( \p_{\nu}\Pom^{\mu}_{\;\;\beta} \right) 
\left( \p_{\delta}\p_{\mu}\Pom^{\nu}_{\;\;\gamma} - \p_{\delta}\p^{\nu}\Pom_{\gamma \mu} \right)
\label{2.5}
\end{eqnarray}
with $\varkappa'$ a dimensionless constant and
$\varkappa''$ a constant of dimension GeV$^{-2}$.

For our fictitious reaction (\ref{2.1}) there is
strict equivalence
\begin{equation}
{\cal L}^{{\rm CS}} \wideestimates 
{\cal L}^{(2,2)} + {\cal L}^{(4,4)}
\label{2.6}
\end{equation}
if the couplings satisfy the relations
\begin{eqnarray}
&&g'_{\Pom \Pom f_{1}} =
-\varkappa'\,\frac{M_{0}^{2}}{k^{2}}
-\varkappa''\,\frac{M_{0}^{2}(k^{2}-2 m^{2})}{2k^{2}} \,,
\nonumber\\
&&g''_{\Pom \Pom f_{1}} =
\varkappa''\,\frac{2 M_{0}^{4}}{k^{2}} \,.
\label{2.7}
\end{eqnarray}

For our CEP reaction (\ref{1.1}) we are dealing with pomerons of mass
squared $t_{1}, t_{2} < 0$ and, in general, $t_{1} \neq t_{2}$.
Then, the equivalence relations (\ref{2.6}), (\ref{2.7}),
will still be approximately true and we confirm this
by explicit numerical studies.

\section{Results for the WA102 experiment}
\vspace{-0.3cm}

Many experimental results for CEP in proton-proton collisions
at a c.m.~energy of $\sqrt{s} = 29.1$~GeV
have been obtained by the WA102 collaboration
in the years 1997--2000.
They worked at the Omega spectrometer at CERN
and they could measure the complete final state:
the central meson plus the outgoing protons.
They obtained for $\sqrt{s} = 29.1$~GeV
the following total cross sections for a cut
on the meson's Feynman variable $|x_{F}| \leqslant 0.2$ (see \cite{Barberis:1998by}):
\begin{eqnarray}
f_{1}(1285):&& \sigma_{\rm exp.} = (6919 \pm 886)\;\mathrm{nb}\,,
\nonumber \\
f_{1}(1420):&& \sigma_{\rm exp.} = (1584 \pm 145)\;\mathrm{nb}\,.
\label{3.1}
\end{eqnarray}
The WA102 collaboration also gave distributions in $t$ and
in $\phi_{pp}$ ($0 \leqslant \phi_{pp} \leqslant \pi$),
the azimuthal angle between the transverse momenta
of the two outgoing protons in the overall c.m. system.

\begin{figure}[!h]
\center
\includegraphics[width=0.46\textwidth]{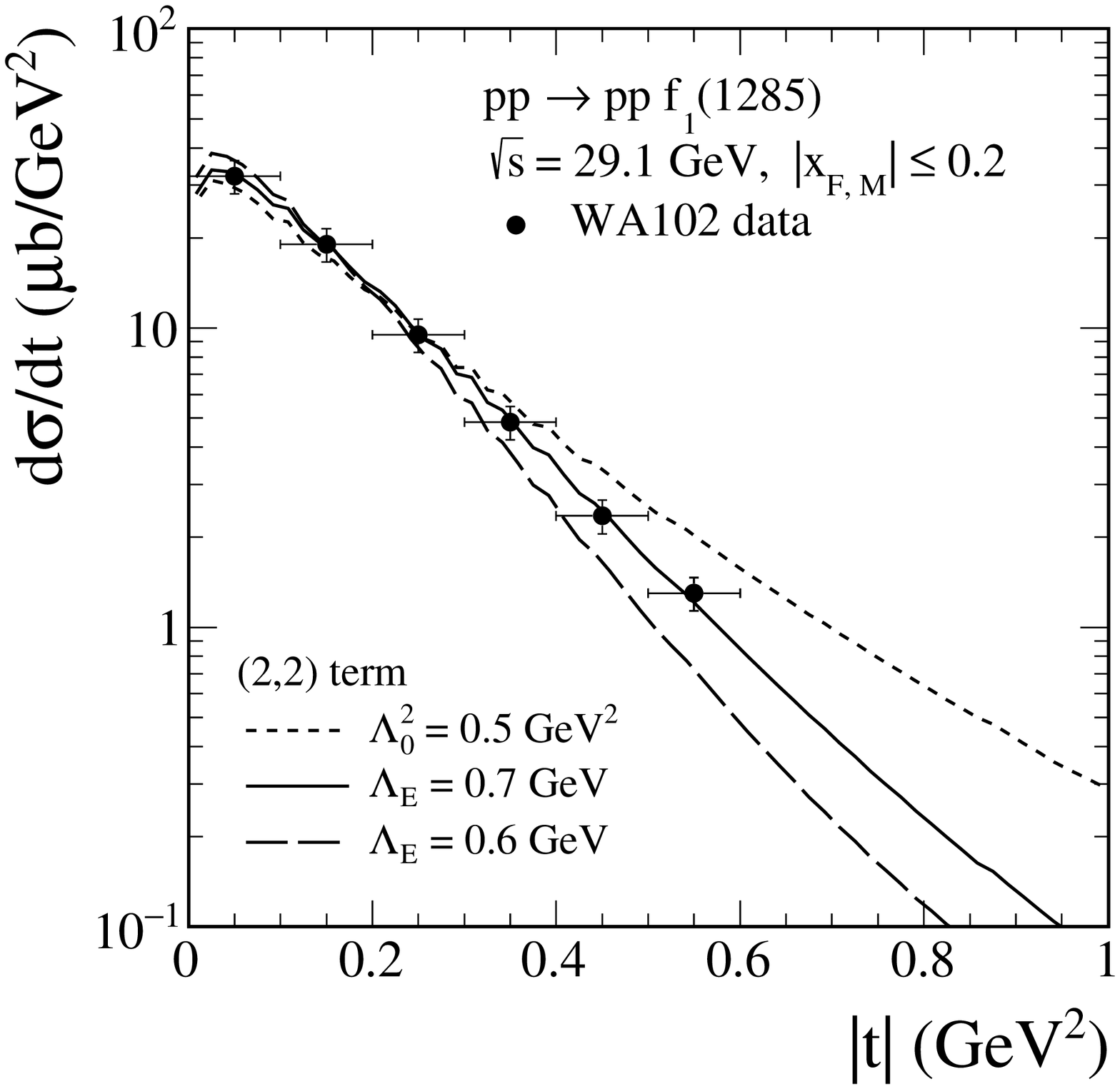}
\includegraphics[width=0.46\textwidth]{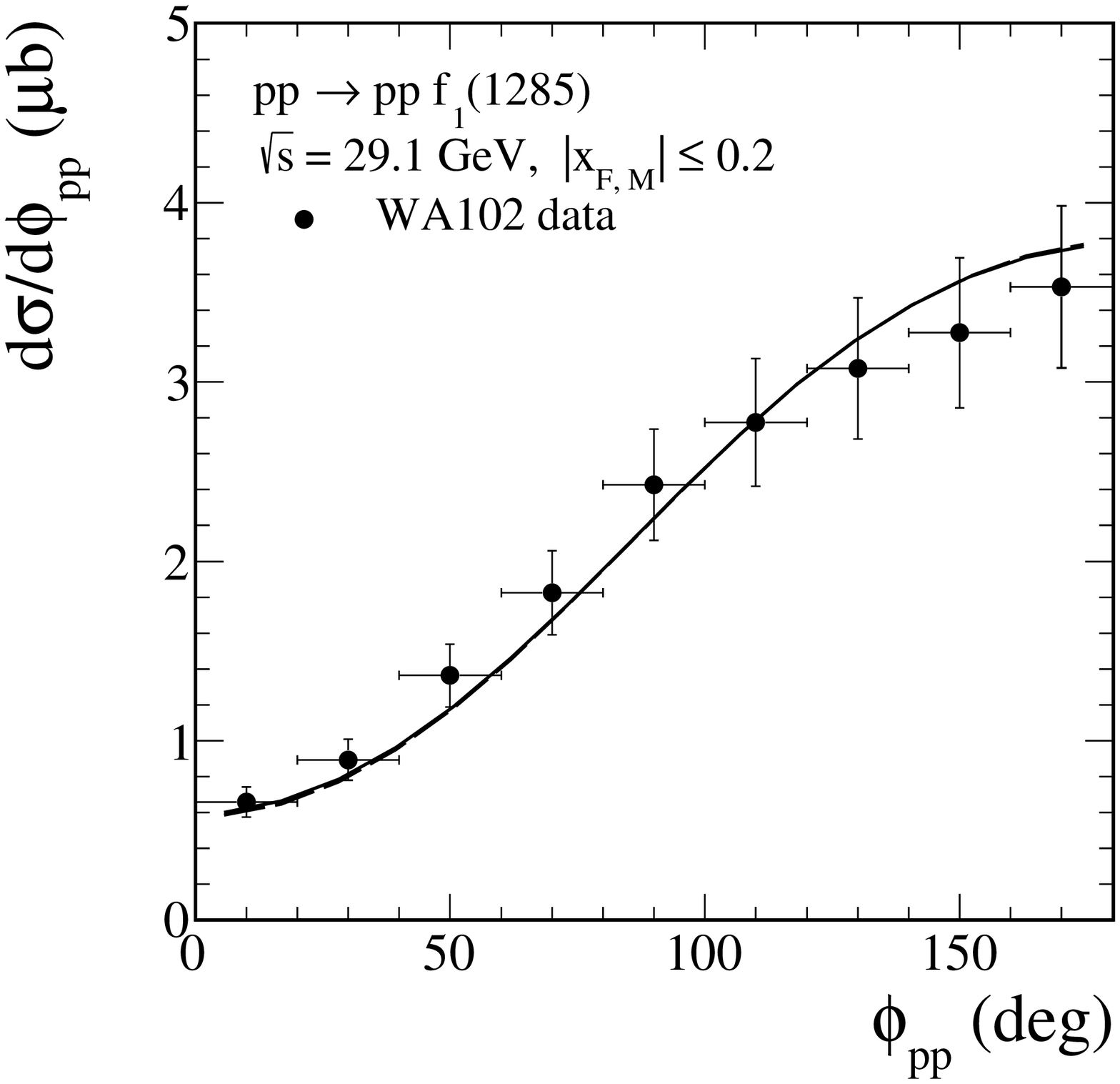}
\caption{Fit to the WA102 data using the $(2,2)$ term only,
$|g'_{\Pom \Pom f_{1}}| = 4.89$; see (\ref{2.3}).}
\label{fig5}
\end{figure}

\begin{figure}[!h]
\center
\includegraphics[width=0.46\textwidth]{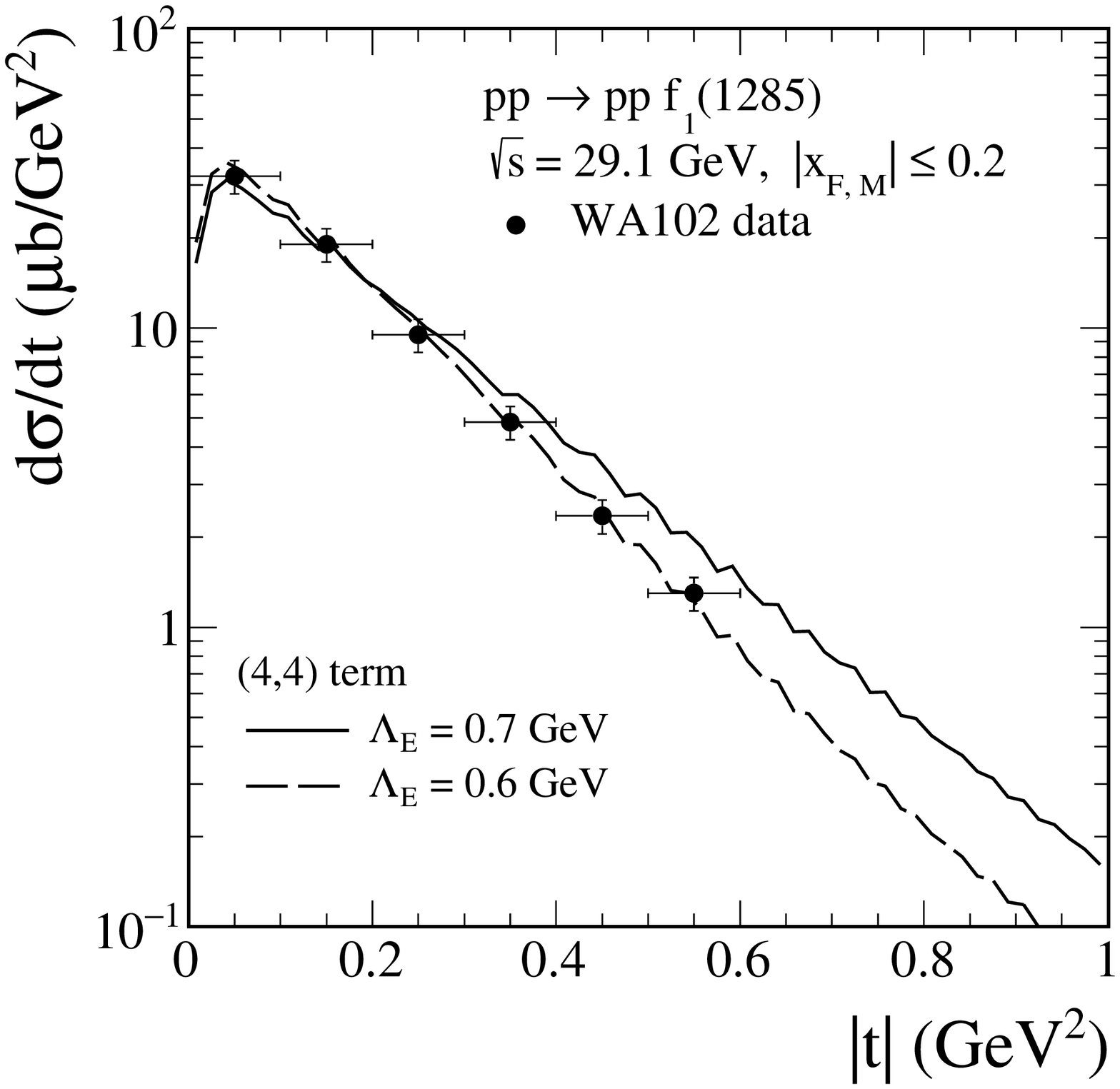}
\includegraphics[width=0.46\textwidth]{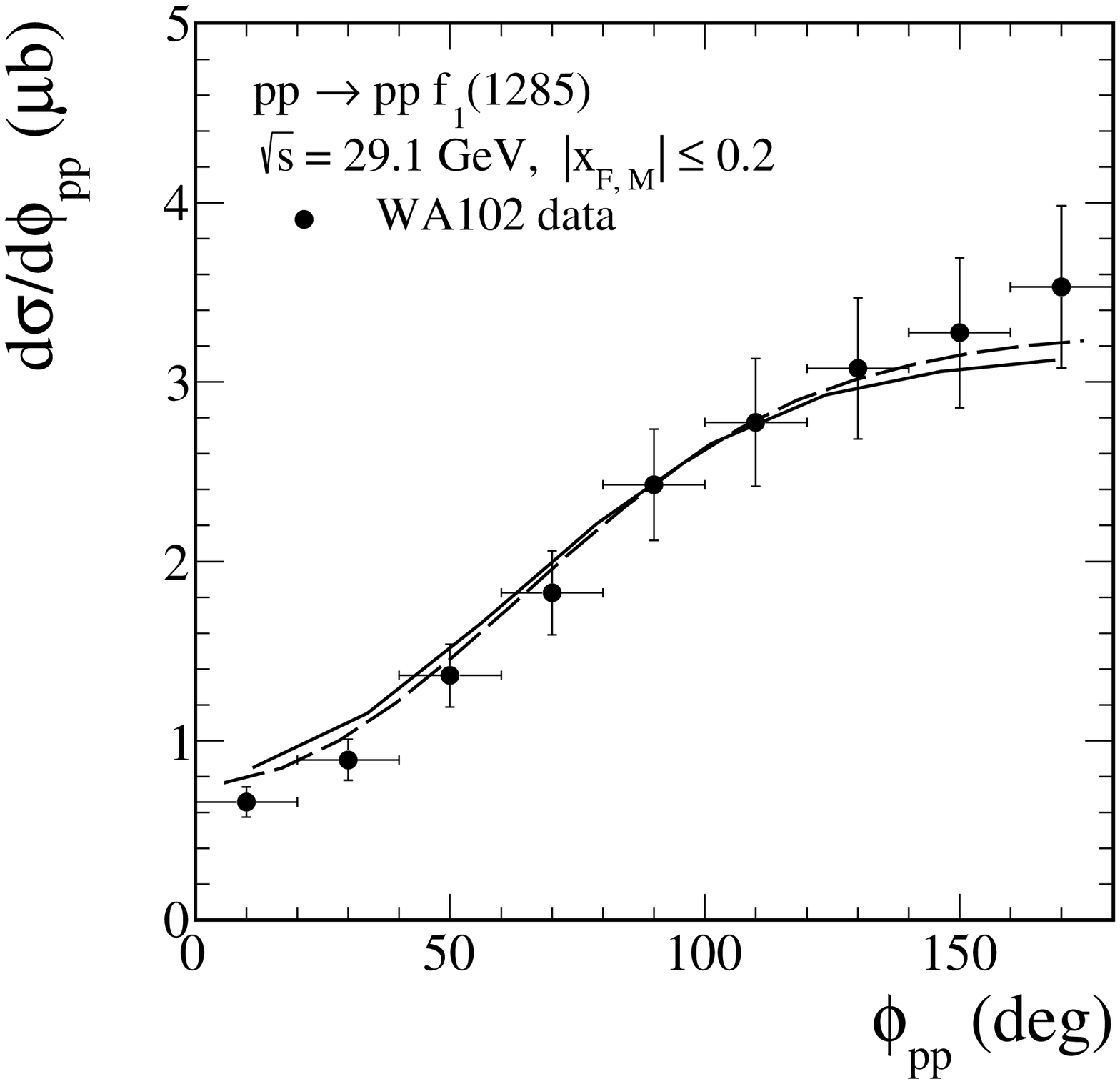}
\caption{Fit to the WA102 data using the $(4,4)$ term only,
$|g''_{\Pom \Pom f_{1}}| = 10.31$; see (\ref{2.4}).}
\label{fig6}
\end{figure}

\begin{figure}[!h]
\center
\includegraphics[width=0.44\textwidth]{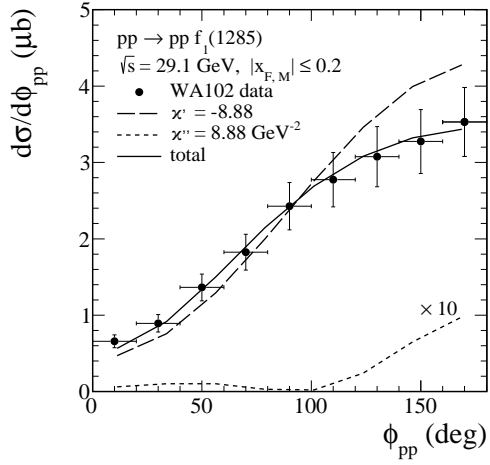}
\caption{Fit to the WA102 data using the Chern-Simons coupling
with $\varkappa'= -8.88$, $\varkappa''/\varkappa' = -1.0$~GeV$^{-2}$;
see (\ref{2.5}).
The $\varkappa''$ contribution has been
enhanced by a factor 10 for better visibility.}
\label{fig7}
\end{figure}

We are assuming that the reaction (\ref{1.1})
is dominated by pomeron exchange (figure~\ref{fig1}) already
at $\sqrt{s} = 29.1$~GeV.
Using this we have calculated in
our tensor-pomeron approach the cross sections,
the $t$ and $\phi_{pp}$ distributions
for $f_{1}(1285)$ and $f_{1}(1420)$ CEP,
and compared our results to the WA102 data.
In figures~\ref{fig5}--\ref{fig7} we show 
some of our results \cite{Lebiedowicz:2020yre}
which include - very important - absorptive corrections.
Here $\Lambda_{E}$ is a form-factor parameter.
We get a reasonable description of the WA102 data
with $\Lambda_{E} = 0.7$~GeV and the following possibilities:
\begin{eqnarray}
(l,S) = (2,2)\;\mathrm{term \; only}:&&
g'_{\Pom \Pom f_{1}} = 4.89\,, \;
g''_{\Pom \Pom f_{1}} = 0;
\label{3.2}\\
(l,S) = (4,4)\;\mathrm{term \; only}:&&
g'_{\Pom \Pom f_{1}} = 0\,, \;
g''_{\Pom \Pom f_{1}} = 10.31;
\label{3.3}\\
\mathrm{CS \; terms}:&&
\varkappa' = -8.88\,, \;
\varkappa''/\varkappa' = -1.0\;\mathrm{GeV}^{-2}\,.
\label{3.4}
\end{eqnarray}


Now we can use our equivalence relation (\ref{2.7})
in order to see to which $(l,S)$ couplings (\ref{3.4})
corresponds.
Replacing in (\ref{2.7}) $m^{2}$ by 
$t_{1} = t_{2} = -0.1$~GeV$^{2}$
and $k^{2}$ by $m_{f_{1}}^{2} = (1282 \;\mathrm{MeV})^{2}$
we get from (\ref{3.4})
\begin{equation}
g'_{\Pom \Pom f_{1}} = 0.42\,, \;
g''_{\Pom \Pom f_{1}} = 10.81\,.
\label{3.5}
\end{equation}
Thus, the CS couplings of (\ref{3.4}) correspond
to a nearly pure $(l,S) = (4,4)$ coupling (\ref{3.3}),
and the corresponding values of $g''_{\Pom \Pom f_{1}}$
of (\ref{3.5}) and (\ref{3.3}) agree to within 5\,\%.

\section{Conclusions}
\vspace{-0.3cm}
\begin{enumerate}
 \item[$\bullet$] 
We have discussed in detail the forms of the $\Pom \Pom f_{1}$
coupling.

 \item[$\bullet$] We obtain a good description of the WA102 data at
$\sqrt{s} = 29.1$~GeV.

 \item[$\bullet$] Our results for higher energies indicate similar distributions
as at the lower energy and cross sections for CEP of
the $f_{1}(1285)$ of
$\sigma \cong 30-140 \;\mathrm{nb}$ 
for the STAR experiment at RHIC
and $\sigma \cong 6-40 \;\mu\mathrm{b}$
for the LHC experiments,
depending on the assumed cuts.

 \item[$\bullet$] Detailed tests of the Sakai-Sugimoto model are possible.

 \item[$\bullet$] Experimental studies of single meson CEP reactions will allow
to extract many pomeron-pomeron-meson coupling parameters.
Their theoretical calculation is a challenging
problem of nonperturbative QCD.

\end{enumerate}

\acknowledgments
\vspace{-0.3cm}

The authors thank the organisers of the ICHEP 2020
conference for making this presentation of our results possible.
This work was partially supported by
the Polish National Science Centre under Grant
No. 2018/31/B/ST2/03537.
J.L. was supported by the Austrian Science Fund FWF,
doctoral program Particles \& Interactions, Project No. W1252-N27.

\end{document}